\documentstyle[prl,aps,epsf,psfig,cite]{revtex}
\input epsf.sty
\begin{document}
\twocolumn[\hsize\textwidth\columnwidth\hsize\csname@twocolumnfalse%
\endcsname
\title{Confinement of electrons in layered metals}
\author{M.A.H. Vozmediano$^{\dag}$,
M. P. L\'opez Sancho$^{\ddag}$ and F. Guinea$^{\ddag}$.}
\address{
$^{\dag}$ Departamento de Matem\'aticas,
Universidad Carlos III de Madrid, Avda. de la Universidad 30,28911 Legan\'es, 
Madrid, Spain. \\                                                             
$^{\ddag}${I}nstituto de Ciencia de Materiales de Madrid,
CSIC, Cantoblanco, E-28049 Madrid, Spain.}
\date{\today}
\maketitle
\begin{abstract}
We analyze the out of plane hopping in models of layered systems where
the in--plane properties deviate from Landau's theory of a Fermi liquid.
We show that the hopping term acquires a non trivial energy dependence,
due to the coupling to in plane excitations,
and can be either relevant or irrelevant at low energies or temperatures.
The latter is always the case if the Fermi level lies close to
a saddle point in the dispersion relation.
\end{abstract}
]
\narrowtext
Layered materials have been object of intensive study
since they present important physics. Unusual properties
are derived from the anisotropy and periodicity along the
axis perpendicular to the planes \cite{quinn}. 
Among the most studied layered materials are the  
high-temperature cuprate superconductors.
These compounds present a strong anisotropy 
and are treated as two-dimensional systems in many approaches.
In the normal state the transport properties within the CuO$_2$
planes are very different from those along the c-axis: electron
motion in the c-direction is incoherent in contrast with the 
metallic behavior of the in-plane electrons as probed by the 
different {$\bf \rho_{c}$} and {$\bf \rho_{ab}$} resistivities
and their different dependence with temperature \cite {fuji, ando}.
Optical conductivity measurements confirm the anomalous c-axis 
properties \cite {review}. 
The relevance of the
nature of the conductance in the
direction perpendicular to the CuO$_2$ planes to the nature
of the superconducting phase has been remarked
on  both theoretical \cite {ander,legg,TL01}
and  experimental \cite {marel} grounds.
The anomalous nature of the out of plane properties in the cuprates,
and in analogy with the one dimensional Luttinger liquid, 
has led to the proposal of the
failure of the conventional Fermi-liquid theory in these compounds\cite{ander}.
An alternative explanation of the 
emergence of incoherent behavior in the out of plane direction has been
proposed in terms of the coupling of the interlayer electronic motion to
charge excitations of the system\cite{TL01}. This approach implicitly 
assumes that electron electron interactions modify the in--plane
electron propagators in a non trivial way, at least at distances
shorter than the elastic mean free path, $l$. The strong angular dependence
of both the scattering rate and the interplane tunneling element
can also lead to the observed anisotropies\cite{IM98}.

Graphite, another layered material, presents a intraplane
hopping much larger than the interplane hybridization. 
The unconventional transport properties of graphite such
as the linear increase with energy of the inverse 
lifetime\cite{Yetal96} (see also\cite{STK01}),
suggest deviations from the conventional Fermi liquid 
behavior, which could be due to strong Coulomb
interactions unscreened because of the lack of states at the
Fermi level \cite{GGV94}.

In the following, we extend the methods used in\cite{TL01}
to the clean limit, $l \rightarrow \infty$, and show that, 
for certain 
models of correlated electrons\cite{GGV94,GGV96} the
interplane hopping between extended states can be a relevant
or an irrelevant variable, in the RG sense, depending on the strength of
the coupling constant.
The scheme used here is based on the
renormalization group analysis 
as applied to models of interacting electrons\cite{S94,P92}. 
For the problem of
interchain hopping
between Luttinger liquids, it recovers well known 
results\cite{W90,KF92,1D}.
 
We first present the method of calculation, and show the results it
gives in one dimension. Then, we apply it to 
two dimensional models which show deviations from FL behavior.
The main physical consequences of our calculation are discussed
in the last section.

{\it The method of calculation.}
In the presence of electron-electron interactions, tunneling processes
are modified by inelastic scattering events. If the excitations of the 
systems (electron-hole pairs, plasmons) are modelled as bosonic
modes, one can write an effective electron-boson hamiltonian of the type:
\begin{equation}
{\cal H}_{e-b} = \sum {\bf t}_{ij} c^\dag_i c_j + \sum \omega_k b^\dag_k
b_k + \sum g_{k , i} c^\dag_i c_i ( b^\dag_k + b_k )
\label{hamil}
\end{equation}
where the ${\bf t}_{ij}$ describe the electronic hopping processes. 
The electron-boson interaction leads to a Franck Condon factor which
reduces the effective tunneling rate. The electron propagators
acquire an anomalous time, or energy, dependence:
\begin{eqnarray}
\langle c^\dag_i ( t ) c_j ( t' )
\rangle &
\sim &\frac{C_{ij}}{t - t'} \times \nonumber \\ &\times &\exp \left\{
\int d \omega \left[ 1 - e^{i \omega ( t - t' )} \right]
\frac{J_{ij} ( \omega )}
{ \omega^2} \right\}
\label{Green}
\end{eqnarray}
where:
\begin{equation}
J_{ij} (  \omega ) = \sum_k | g_{i k} g_{j k} |^2
\delta ( \omega - \omega_k )
\end{equation}
This expression can be generalized, taking into account the
spatial structure of the coupling to:
\begin{eqnarray}
\langle \Psi^{\dag}_i ( t ) \Psi_j ( t' )
\rangle &
\sim &\frac{C_{ij}}{t - t'} \times \nonumber \\ &\times &\exp \left\{
\int_{{\bf \vec{k}} }
 \int_{ \Omega} \int_{ \Omega }
d^2 {\bf \vec{r}} d^2 {\bf \vec{r}'}
d^2 {\bf \vec{k}} e^{i {\bf \vec{k}}
( {\bf \vec{r} - \vec{r}'} )} \right. \nonumber \\ & & \left.
\int d \omega \left[ 1 - e^{i \omega ( t - t' )} \right]
\frac{V_{eff} ( {\bf \vec{k}} , \omega )}
{ \omega^2} \right\}
\label{propagator}
\end{eqnarray}
where $\Omega$ is the region of overlap of the wavefunctions
$\Psi_i ( {\bf \vec{r}} )$ and 
$\Psi_j ( {\bf \vec{r}} )$.
This expression, which can be seen as the exponential of the leading
frequency dependent self--energy correction to the electron propagator,
has been extensively used in studies of tunneling in zero dimensional 
systems (single electron transistors) which show Coulomb 
blockade\cite{SET}, one dimensional conductors\cite{SNW95}, and
disordered systems in arbitrary dimensions\cite{RG01}.

The effective interaction can, in turn, be written in terms
of the response function as:
\begin{equation}
V_{eff} ({\vec {\bf k}} , \omega ) = V^2 ( {\bf k} )
{\rm Im} \chi ({\vec {\bf k}} , \omega )
\label{veff}
\end{equation}

In a Fermi liquid, we have
$V_{eff} ({\vec {\bf k}} , \omega ) \approx \alpha ( {\vec{\bf k}} )
\vert \omega \vert$ for $\omega \ll E_F$, where $E_F$ is the Fermi energy,
so that:
\begin{eqnarray}
\lim_{(t-t') \rightarrow \infty} \langle &\Psi^{\dag} ( t ) \Psi ( t' )
\rangle 
&\sim \frac{1}{( t - t' )^{1 + \bar{\alpha}}} \nonumber \\
\bar{\alpha} &=  
&\int_{{\bf \vec{k}}}
\int_{\Omega} \int_{\Omega}
d^2 {\bf \vec{r}} d^2 {\bf \vec{r}}' 
d^2 {\bf \vec{k}} e^{i {\bf \vec{k}}
( {\bf \vec{r} - \vec{r}'} )}  \alpha ( {\bf \vec{k}} )
\label{alpha}
\end{eqnarray}
The parameter $\bar{\alpha}$ gives the correction to the
scaling properties of the Green's functions.
It is easy to show that, in an isotropic Fermi liquid in D dimensions,
$\lim_{l \rightarrow \infty}
\bar{\alpha} \propto l^{1 - D}$, where $l$ is the linear
dimension of the (localized) electronic wavefunctions $\Psi
( {\bf \vec{r}} )$. This result is due to
the dependence on ${\bf \vec{k}}$ of the response
functions, as ${\rm Im} \chi ( {\bf \vec{k}} ,
\omega ) \sim | \omega | / ( k_F^{D-1} | {\bf \vec{k}} | )$.
Thus, for $D > 1$, we recover coherent tunneling
in the limit of delocalized wavefunctions.

In one dimension, one can use the non interacting 
expression for ${\rm Im} \chi_0 ( {\bf \vec{k}} ,
\omega )$, to obtain:
\begin{eqnarray} 
& &\int
d^2 {\bf \vec{r}} d^2 {\bf \vec{r}'}
d^2 {\bf \vec{k}} e^{i {\bf \vec{k}}
( {\bf \vec{r} - \vec{r}'} )}
\int d \omega \left[ 1 - e^{i \omega ( t - t' )} \right]
\frac{V_{eff} ( {\bf \vec{k}} , \omega )}
{ \omega^2} \nonumber \\ 
&\propto &\left( \frac{U}{E_F} \right)^2 \times \left\{
\begin{array}{lr} 0 &t - t' \ll l / v_F \\
\log [ v_F ( t - t' ) / l ] &t - t' \gg l / v_F \end{array}
\right.
\label{1D}
\end{eqnarray}
where we have assumed a short range interaction, $U$.
Hence, the Green's functions have a
non trivial power dependence on time, even in the
$l \rightarrow \infty$ limit, in agreement with 
well known results for Luttinger liquids\cite{SNW95}.
In order to obtain the energy dependence of the effective
tunneling between ${\bf \vec{k}}$ states near the Fermi surface, one
needs to perform an additional integration over $d {\bf \vec{r}}$.
In general, near a scale invariant fixed point,
$\omega \propto | {\bf \vec{k}} |^z$, and for a 1D conductor
one knows that $z=1$. Hence, ${\rm Im} G ( \omega , k_F )
\propto \omega^{-z + \bar{\alpha}}
\sim \omega^{-1 + \bar{\alpha}}$. 
The flow of the hopping terms under a Renormalization Group
scaling of the cutoff is \cite{1D}:
\begin{equation}
\Lambda \frac{\partial ( {\bf t} / \Lambda )}{\partial \Lambda} =
\left\{ \begin{array}{lr} - \bar{\alpha} &{\rm localized \, \, \, hopping} \\
1 - \bar{\alpha} &{\rm extended \, \, \, hopping} \end{array} \right.
\label{scaling}
\end{equation}
where ${\bf t}$ denotes a hopping term, between localized or extended states.
In the latter case,
the hopping becomes
an irrelevant variable\cite{W90} for $\bar{\alpha} > 1$.

{\it Graphene planes.} 
The simplest two dimensional model for interacting electrons where
it can be rigourously shown that the couplings acquire logarithmic
corrections in perturbation theory is a system of Dirac fermions
($\epsilon_k = v_F | {\bf \vec{k} |}$), with Coulomb, $1 / |
{\bf \vec{r} - \vec{r}'} |$, interaction. This model can be used
to describe isolated graphene planes\cite{GGV94}, and can 
help to understand the anomalous behavior of graphite observed
in recent experiments\cite{Yetal96,STK01}. 

In order to apply the procedure outlined in the previous section, one 
needs the Fourier transform of the interaction, 
$e^2 / ( \epsilon_0
| {\bf \vec{k}} |)$, where $e$ is the electronic charge, and 
$\epsilon_0$ is the dielectric constant, and the susceptibility
of the electron gas. For a single graphene plane, this quantity
is:
\begin{equation}
\chi_0 ({\bf \vec{k}}, \omega ) = \frac{1}{8} \frac{ | {\bf
\vec{k}}|^2} {\sqrt{v_F^2 |{\bf \vec{k}}|^2 - \omega^2 }}
\label{chigraphite}
\end{equation}
These expressions need to be inserted 
in equations (\ref{veff}) and (\ref{propagator}).
Alternatively, we can use the RPA, and include the effects
of interplane screening:
\begin{equation} 
\begin{array}{ll}
\chi_{RPA} ({\vec{\bf k}}, \omega  ) &= \\
& \frac{\sinh ( | {\bf \vec{k}} | d )}
{\sqrt{ \left[ \cosh ( | {\vec{\bf k}} | d ) + \frac{2 \pi e^2}
{ \epsilon_0 | {\bf k} |}
\sinh ( | {\vec{\bf k}} | d )
\chi_0 ( {\vec{\bf k}} , \omega ) \right]^2 - 1 }} \end{array} 
\label{RPA}
\end{equation}
where $d$ is the interplane spacing. The imaginary part, ${\rm Im} \chi_0
( {\bf \vec{k}} , \omega )$, is different from zero if $\omega >
v_F | {\bf \vec{k}} |$.

For simplicity, we consider the expression in eq.(\ref{chigraphite}),
as it allows us to obtain analytical expressions. We cut off the spatial
integrals at a scale, $l$, of the order of the electronic
wavefunctions involved in the tunneling. We obtain an expression
similar to that in eq.(\ref{1D}) except that the prefactor
$( U / E_F )^2$ is replaced by $e^4 / ( \epsilon_0 v_F )^2$.
Thus, the propagators acquire an anomalous dimension. As in 1D,
the value of the exponent $z$ which relates length and time scales
is $z=1$. The scaling of the hoppings now are:
\begin{equation}
\Lambda \frac{\partial ( {\bf t} / \Lambda )}{\partial \Lambda} =
\left\{ \begin{array}{lr} - 
1 - \bar{\alpha} &{\rm localized \, \, \, hopping} \\
1 - \bar{\alpha} &{\rm extended \, \, \, hopping} \end{array} \right.
\label{scaling_graphite}
\end{equation}
The extra constant in the first equation with respect to eq. (\ref{scaling})
reflects the vanishing of the density of states at the Fermi level for
two dimensional electrons with a Dirac dispersion relation.

In graphite, the dimensionless coupling constant,  $e^2 / v_F$,
is of order unity. Under renormalization, it flows towards
zero\cite{GGV94}.
Thus, interplane tunneling is a relevant variable,
although, in a dirty system with a finite
mean free path, interplane tunneling can become irrelevant\cite{GGV01}.

{\it Saddle point in the density of states.}
The Fermi surface of most hole-doped cuprates is close to a Van Hove
singularity \cite{vhexp}.  The possible relevance of this fact to the
superconducting transition as well as to the anomalous behavior of the
normal state was put forward in the early times of the cuprates 
and gave rise to the socalled Van Hove scenario  \cite{vhscenario}.

We shall apply the  mechanism described in section  II
to study the interlayer hopping of two electronic systems described by the
{\bf t--t'} Hubbard model and whose Fermi surface lies close to a Van Hove 
singularity.

The {\bf t-t'}-Hubbard
model has the dispersion relation
\begin{eqnarray}
\varepsilon({\vec{\bf k}})& =& -2{\bf t}\;[ \cos(k_x a)+\cos(k_y a)] \nonumber \\
&-&2{\bf t'}\cos(k_x a)\cos(k_y a)
\;\;\;, 
\label{disp0}
\end{eqnarray}
This dispersion relation has two inequivalent saddle points at 
A $(\pi,0)$ and  B $(0,\pi)$.
The Van Hove model in its simplest 
formulation is obtained by 
assuming that for fillings such that
the Fermi line lies close to the singularities, the majority of
states participating in the interactions will come from regions in the
vicinity of the saddle points\cite{tperp}. 
Taylor expanding eq. (\ref{disp0}) around
the two points gives the effective relation
\begin{equation}
\varepsilon_{A,B} ({\vec {\bf k}} ) \approx \mp ( {\bf t} \mp 2 {\bf t'} ) k_x^2 
\pm ( {\bf t} \pm 2 {\bf t'} ) k_y^2  \;\;\;,
\label{disp}
\end{equation}
which leads to the effective hamiltonian:
\begin{eqnarray}
{\cal H} &= &\sum_{i=1,2;k,s}^{\epsilon_{i,k} < \Lambda_0}
 \epsilon_{i,k} c^\dag_{k,i,s}
c_{i,k,s} \nonumber \\ &+ &\sum u_{i,i';s,s'} c^\dag_{i,k,s} c_{i',k',s'}
c^{\dag}_{i'',k'',s''} c_{i''',k''',s''' }\;,
\end{eqnarray}
where $\Lambda_0$ is a high energy cutoff which sets the limit of 
validity of the effective description.
The particle--hole susceptibility has been  computed in \cite{GGV96}:
%
%
\begin{equation}
{\rm Im} \: \chi ({\vec{\bf k}}, \omega) = 
\frac{1}{4 \pi\varepsilon ({\vec{\bf k}})} \left(\left| \omega+\varepsilon 
({\vec{\bf k}})
\right| - \left| \omega-\varepsilon ({\vec{\bf k}}) \right|\right)\;,
\end{equation}
where $\varepsilon({\vec{\bf k}})$ is the dispersion relation (\ref{disp}).

The long time dependence of the Green's function is determined
by the low energy behavior of $\chi$: 
$$\lim_{\omega w\to 0}{\rm Im} \: \chi ({\vec{\bf k}}, \omega) \sim 
\frac{\omega}{\varepsilon ({\vec{\bf k}})}$$
Inserting this expression in eqs.(\ref{propagator}) and
(\ref{veff}), 
we can see that, irrespective of the details
of the interaction,  in the presence of
a Van Hove singularity the exponent $\bar{\alpha}$ in the
time dependence of the Green's function goes as:
$$\lim_{\Omega \rightarrow
\infty} \bar{\alpha} \propto \log ( l )\;\;,$$
where $l$, as before, is the length scale which characterizes
the wavefunction of the tunneling electron, and $\Omega \propto
l^2$ is the size of the integration region in eq.(\ref{propagator}).

The details of the anomalous dimension of the propagator in this case
depend on the nature of the interactions which determine
$V_{eff}({\bf \vec{k}},\omega)=V^2({\bf \vec{k}})
{\rm Im} \chi ( {\bf\vec{k}} , \omega )\;$.
To make contact with previous works \cite{TL01},
we have computed the parameter $\bar{\alpha}$ for 
different possible interactions: 

a) Unscreened Coulomb potential,
$ V({\bf \vec{k}})=\frac{2\pi e^2}{\epsilon_0 \vert 
{\bf \vec{k}}\vert}$. 
Due to the highly singular interaction, the $\omega$
dependence of the effective potential is not linear and $\bar{\alpha}$
is not well defined. The effective potential eq. (\ref{veff}) is
computed to be:
$$V_{eff}({\bf \vec{k}},\omega)=\frac{ e^4}{\epsilon_0^2 {\bf t}}
\left[ 1+\log\left(\frac{\omega}{\omega_0}\right)
\right]$$
where $\omega_0 = | E_F - E_{VH} |$ is a low-energy 
cutoff, required to avoid divergences in the integrals. 
The Fermi energy is $E_F$ and the position of the saddle point is
at $E_{VH}$.

b) 
Screened interaction of the  Hubbard type, $V({\bf \vec{k}})=U a^2$,
where $a$ is the lattice unit.
$$\bar{\alpha}=\frac{4U^2}{\pi {\bf t}^2}\frac{1}{(2\pi)^2}K_M
\left[1+\frac{1}{2}
\log\left(\frac{\Lambda}{\omega}\right)\right]\;\;,
$$
where $\Lambda$ is a high-energy cutoff, and 
$$K_M=\frac{1}{2}\log \left\vert\frac{k_x+k_y}{k_x-k_y}\right\vert\;,$$
where $k_x , k_y \sim \sqrt{\omega_0}$.

c) Thomas--Fermi screened potential, 
$$ V({\bf \vec{k}})^2=\frac{(2\pi e^2)^2}{\epsilon_0^2(k^2 + k_{FT}^2)}\;\;\;,\;\;\;
\bar{\alpha}\sim\frac{2e^4 k_{FT}^2}
{\epsilon_0\pi {\bf t}^2}\log^2(k_{FT} / k)\;,$$
where $k_{FT}$ is the Fermi-Thomas wavevector,
and, as before, $k \sim \sqrt{\omega_0}$. This is an
intermediate situation  between the two previous cases.

Finally, in the case of a layered system, we can use the RPA summation,
including interlayer interactions, eq.(\ref{RPA}).
In this case, $z = 2$, and the scaling equations for the
hoppings are the same as in eq.(\ref{scaling}).

\begin{figure}
\begin{center}
\mbox{\psfig{file=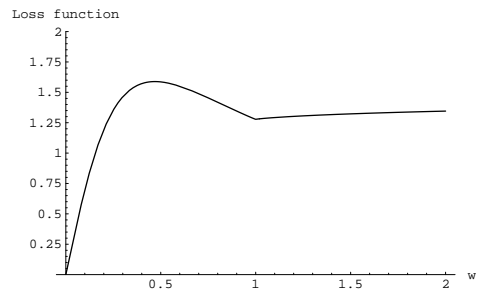,width=.35\textwidth}}
\end{center}
\caption{Effective potential eq. (\ref{veff}) as a function
of the energy for fixed ${\bf \vec{k}}$ for a system of Van Hove
layers coupled by Coulomb interaction .} 
\label{loss}
\end{figure}

A numerical computation provides the effective potential 
shown in fig. [\ref{loss}].

Fig. [\ref{loss}]  compares well with the experimental plots
of the loss function given in \cite{curvaexp},  what reveals that
the Van Hove model is also compatible with transport experiments.
The results for the parameter
$\bar{\alpha}$ extracted from the numerical computation
are in qualitative agreement with the analytical
expressions given in cases b) and c) above.

In all cases $\bar{\alpha}$ diverges as $E_F 
\rightarrow E_{VH}$. Thus, interlayer hopping is an 
irrelevant variable, and scales towards zero as the
temperature or frequency is decreased. This additional
logarithmic dependence can be seen as a manifestation
of the $\log^2$ divergences
which arise in the treatment of this model\cite{GGV96}.
Note that, as in the graphene case, the coupling
constants are also energy dependent, and grow at
low energies, suppressing even further the interlayer
tunneling.

{\it Conclusions.}
We have discussed the supression of interlayer tunneling by
inelastic processes in two dimensional systems in the clean limit.
Our results suggest that, when perturbation theory
for the in--plane interactions leads to logarithmic
divergences,
the out of plane tunneling acquires
a non trivial energy dependence. This anomalous scaling
of the interlayer hopping can make it irrelevant, at
low energies, if the in--plane interactions are sufficiently
strong. This is always the case if the Fermi level
of the interacting electrons lies at a van Hove 
singularity (note that the Fermi level can, in certain
circumstances, be pinned to the singularity\cite{GGV96}). 
Thus, we have shown that insulating behavior in the
out of plane direction is not incompatible with
gapless
in--plane properties.

{\it Acknowledgements}.
We are thankful to R. Markiewicz for a careful reading of
the manuscript.
The financial support of the CICyT (Spain), through
grants no. PB96-0875, 
1FD97-1358 and BFM2000-1107,  
is gratefully acknowledged.



\begin{references}
\bibitem{tperp}
Note that the momentum dependence of the out of plane hopping in the 
cuprates
tends to suppress tunneling from points away from the singularities.
\end{references}
\end{document}